\documentclass[11pt]{article}
 
\usepackage{graphicx,epsf, epsfig, amssymb, amsmath, amsfonts, mathrsfs}
\usepackage{longtable}
\usepackage{bm}
\usepackage{color}
\usepackage{enumitem}
\usepackage{float}
\usepackage{caption}
\usepackage{subcaption}
\usepackage{upgreek}
\usepackage{cite}
\usepackage{mathtools}
\usepackage{titlesec}
\usepackage{lipsum}
\usepackage{xcolor}
\usepackage[section]{placeins}
\usepackage[colorlinks=true,linktocpage=true,linkcolor=blue,citecolor=blue]{hyperref}
\usepackage{tensor}
\usepackage{multirow}
\usepackage[utf8]{inputenc}
\usepackage{authblk}
\usepackage{array}
\usepackage{breqn}

\newcommand{\be}{\begin{equation}}
\newcommand{\ee}{\end{equation}}
\newcommand{\bea}{\begin{eqnarray}}
\newcommand{\eea}{\end{eqnarray}}

\newcommand{\dmoff}[1]{}

\newcommand\blfootnote[1]{%
  \begingroup
  \renewcommand\thefootnote{}\footnote{#1}%
  \addtocounter{footnote}{-1}%
  \endgroup
}
\usepackage{comment}
\newcommand{\customrule}{\vspace{0.5cm}\noindent\textcolor{gray}{\rule{\textwidth}{0.5pt}}\vspace{0.5cm}}
 
\usepackage
[
a4paper,
left=1.5cm,
right=1.5cm,
top=3cm,
bottom=4cm,
]{geometry}

\numberwithin{equation}{section} 

\titleclass{\subsubsubsection}{straight}[\subsubsection]
\newcounter{subsubsubsection}[subsubsection]
\renewcommand\thesubsubsubsection{\thesubsubsection.\arabic{subsubsubsection}}
 
\titleformat{\subsubsubsection}[block]
  {\normalfont\normalsize\bfseries}{\thesubsubsubsection}{1em}{}
\titlespacing*{\subsubsubsection}
  {0pt}{3.25ex plus 1ex minus .2ex}{1em}
 
\makeatletter
\def\toclevel@subsubsubsection{4}
\def\l@subsubsubsection{\@dottedtocline{4}{7em}{4em}}
\makeatother
 
\author[1,2,3]{Diego Henrique Carvalho dos Santos}
\author[3,4]{José André Lourenço}
\author[2,3,5]{Davi C. Rodrigues}
\author[2,3]{Etevaldo dos Santos Costa Filho}
 
\affil[1]{\footnotesize Instituto Federal do Espírito Santo, Rod. Gov. José Henrique Sette 184, Cariacica, ES, Brazil}
\affil[2]{\footnotesize Programa de Pós-Graduação em Física, Universidade Federal do Espírito Santo, Av.~Fernando Ferrari 514, Vit\'{o}ria, ES, Brazil}
\affil[3]{\footnotesize Núcleo Cosmo-Ufes, Universidade Federal do Espírito Santo,  Av.~Fernando Ferrari 514, Vit\'{o}ria, ES, Brazil}
\affil[4]{\footnotesize Departamento de Ciências Naturais, Universidade Federal do Espírito Santos, 29932-540, Campus São Mateus, ES, Brazil}
\affil[5]{\footnotesize Departamento de Física, Universidade Federal do Espírito Santo,  Av.~Fernando Ferrari 514, Vit\'{o}ria, ES, Brazil}

\begin{document}
	\title{\bf Test-particle dynamics in a noncommutative \\ deformation of Einstein--Rosen waves}
	\date{}                    
 
	\maketitle
 
	\blfootnote{E-mails: {\tt diego.santos2@ifes.edu.br}, {\tt jose.lourenco@ufes.br}, {\tt davi.rodrigues@ufes.br}, {\tt etevaldo.s.costa@ufes.br}}
 
	\begin{abstract}
	We investigate phenomenological effects induced by a noncommutative deformation of Einstein--Rosen gravitational-waves within the framework of the Riemannian geometry of noncommutative surfaces developed by Chaichian et al. Considering a noncommutative structure between the radial and axial coordinates, we study perturbatively the motion of nonrelativistic test particles in the corresponding deformed geometry. We show that the deformation generates a coupling between the radial and longitudinal sectors, while the radial dynamics remains unaffected by the noncommutative deformation at the perturbative order considered. The induced longitudinal response is governed by a spectrally weighted functional of the gravitational-wave profile, implying that longitudinal observables depend on higher-order spectral moments than their commutative radial counterparts. As a consequence, noncommutative effects exhibit enhanced sensitivity to the ultraviolet structure of the gravitational configuration. We analyze separately monochromatic Einstein--Rosen waves and the localized Weber--Wheeler pulse. In the monochromatic case, the noncommutative correction induces a quasi-periodic longitudinal motion, whereas for the Weber--Wheeler pulse it produces an impulsive longitudinal response analogous to a gravitational velocity memory effect. In addition, we derive an explicit analytical expression in commutative General Relativity for the residual radial velocity generated by the Weber--Wheeler pulse at large radial distances in the perturbative regime.
	\end{abstract}
	\vfill
 

	\newpage
	\customrule
	\tableofcontents

\customrule


\section{Introduction}

Noncommutative geometry arises from the idea that the algebra of functions on spacetime may become noncommutative at sufficiently short distances. Early proposals along these lines date back to Heisenberg, who suggested in a 1930 letter to Peierls that noncommuting coordinates could provide a natural mechanism to regularize ultraviolet divergences in quantum electrodynamics \cite{Jackiw2002,Snyder1947,Madore2000}. Although this idea lost prominence after the development of renormalization techniques, interest in noncommutative spaces reemerged several decades later, initially within mathematics and subsequently in high-energy physics \cite{Connes1994,ConnesDouglasSchwarz1998}. A major development in this direction was the work of Seiberg and Witten \cite{seibergwitten1999}, showing that, in a suitable limit, open-string dynamics in background fields is described by gauge theories formulated on noncommutative spaces. Since then, noncommutative structures have appeared in a broad range of contexts, including gauge theories, quantum field theory, condensed matter systems, and gravity \cite{DouglasNekrasov2001,Susskind2001}.

The possibility that spacetime itself acquires a noncommutative structure at the Planck scale has motivated extensive efforts toward formulating noncommutative versions of General Relativity \cite{Moffat2000,Nicolini2005}. A central physical motivation is the expectation that the classical description of spacetime as a smooth pseudo-Riemannian manifold should cease to be valid in the ultraviolet regime. In particular, arguments combining quantum mechanics and General Relativity suggest the emergence of uncertainty relations involving spacetime coordinates \cite{Doplicher1995}. Reviews discussing these ideas in detail may be found in \cite{Hossenfelder2013,Szabo:2006wx}. From this perspective, noncommutative geometry may provide an effective description of quantum gravitational fluctuations of spacetime itself.

Among the various approaches to noncommutative gravity, some formulations are based on direct star-product deformations of the Einstein--Hilbert action \cite{Aschieri:2005yw}, while others rely on gauge-theoretic constructions via the Seiberg--Witten map \cite{Chamseddine:2000si,Chamseddine:2003we}. Algebraic approaches involving enveloping algebras and noncommutative frame formalisms have also been investigated \cite{Calmet:2005qm,Buric:2008th}. Comprehensive reviews can be found in \cite{Szabo:2006wx,MullerHoissen2008}.

In the present work, we adopt instead the geometric framework developed in \cite{Chaichian:2006ht}, where noncommutativity acts at the level of the embedding geometry itself rather than through a direct deformation of the gravitational action. In this construction, the spacetime metric emerges from a noncommutative deformation of an isometric embedding into a higher-dimensional pseudo-Euclidean space. The formalism generalizes the classical geometry of embedded manifolds and relies on the generalized Nash embedding theorem for pseudo-Riemannian manifolds established independently by Clarke \cite{Clarke1970} and Greene \cite{Greene1970Isometric}. Within this approach, noncommutative corrections are implemented directly at the geometric level through the deformation of the induced metric and connection. The formalism also admits an equivalent algebraic formulation in terms of projective modules \cite{Zhang:2008vk}, placing the embedding construction within the standard algebraic framework of noncommutative differential geometry developed by Connes \cite{Connes1994}. Moreover, its computational simplicity allows explicit calculations to be carried out with essentially the same level of complexity as in ordinary Riemannian geometry \cite{Wang:2008ut}. These features make it particularly suitable for the present investigation of phenomenological consequences of noncommutative gravity.

This framework has been applied to several gravitational backgrounds. Noncommutative deformations of Schwarzschild and Schwarzschild--de Sitter spacetimes were studied in \cite{Wang:2008ut}, while noncommutative plane-fronted gravitational-waves were investigated in \cite{Wang:2009ju}, where exact solutions of a noncommutative version of Einstein's equations were obtained. Applications to gravitational collapse and Oppenheimer--Snyder-type solutions \cite{Oppenheimer1939} were considered in \cite{Sun:2010nas}. Additional applications include deformations of anti-de Sitter spacetime \cite{Momeni:2015laa} and Reissner--Nordström geometries \cite{SotoCampos2018}. These applications, however, have focused primarily on the construction of the deformed geometries themselves, including the construction of exact vacuum solutions \cite{Wang:2009ju} or the resulting singularity structure \cite{Sun:2010nas}. By contrast, the dynamical consequences of such deformations for physical probes and the corresponding effective observables remain comparatively unexplored.

Related dynamical and observational questions have been addressed in other noncommutative settings. In the gravitational-wave context, a geometrically deformed pp-wave background was studied within the noncommutative frame formalism in \cite{Camacho2010}, where the corresponding Jacobi equation was used to analyze effects on polarization. Other approaches keep the gravitational background or its radiative description essentially commutative while introducing noncommutative effects through the matter or particle sector. For instance, noncommutative corrections to compact-binary dynamics and gravitational-wave phases were investigated in \cite{Kobakhidze2016,Jenks2020}, while the response of a bar detector to linearized gravitational waves in noncommutative quantum mechanics was studied in \cite{Gangopadhyay2019}. Test-particle motion and gravitational-wave signatures have also been explored in noncommutative-inspired black-hole geometries, where noncommutativity is modeled through an effective smeared matter distribution rather than through an embedding-based deformation of the spacetime geometry \cite{Ahmed2026}. These works provide complementary realizations of noncommutative effects on physical probes, but differ from the present analysis in the geometric construction, in the sector where noncommutativity is introduced, or in both.

Here we address this gap by investigating the phenomenological consequences of noncommutativity in Einstein--Rosen gravitational-waves. Einstein--Rosen spacetimes are exact, cylindrically symmetric vacuum solutions of the Einstein equations \cite{Griffiths:2009dfa}. Although they are not astrophysically realistic because of their infinite extent along the symmetry axis, they provide an analytically tractable setting for studying leading nonlinear corrections beyond the linearized regime \cite{Chen:2020ktv}. Their explicit structure also makes it possible to derive closed-form expressions for noncommutative corrections. This feature is especially relevant to the embedding-based formalism adopted here, since the deformation is implemented through an isometric embedding rather than by directly modifying the field equations. The Rosen embedding of the Einstein--Rosen spacetime \cite{ROSEN1965} therefore provides a concrete radiative vacuum background in which the noncommutative construction can be carried out while retaining analytical control over the underlying classical geometry.

Restricting ourselves to space-space noncommutativity, symmetry considerations naturally lead to two minimal nontrivial deformations. In the present work, we focus on the one involving the radial and axial coordinates. We then study the motion of nonrelativistic test particles in the deformed geometry. A central feature of the resulting structure is that noncommutative effects do not arise as an independent sector. Instead, the observable corrections are triggered by the gravitational-wave background itself. More precisely, within the perturbative regime considered here, noncommutative contributions appear only through mixed terms involving both the gravitational-wave amplitude and the noncommutativity parameter. The induced longitudinal response is therefore entirely driven by the gravitational dynamics encoded in the Einstein--Rosen wave.

We show that the deformation generates a coupling between the radial and longitudinal dynamics while leaving the radial sector unaffected by the noncommutative deformation at the perturbative order considered. The noncommutative longitudinal response probes higher spectral moments of the gravitational-wave profile than those governing the radial motion, making the noncommutative sector intrinsically more sensitive to ultraviolet components of the spectrum.

We analyze monochromatic Einstein--Rosen waves and Weber--Wheeler pulses. Noncommutativity induces a quasi-periodic longitudinal motion in the monochromatic case and an impulsive response, analogous to a gravitational velocity memory effect, in the pulse case. Within commutative General Relativity, we also derive an explicit perturbative expression for the asymptotic radial velocity generated by the Weber--Wheeler pulse. This second-order effect is associated with the energy carried by the gravitational-wave and remains nonzero even in the time-symmetric ingoing--outgoing configuration.

This paper is organized as follows. In Sec.~\ref{sec:RGNS} we review the noncommutative geometric framework of \cite{Chaichian:2006ht}, focusing on its operational aspects and on the role of isometric embeddings. In Sec.~\ref{sec:NCERW} we construct the noncommutative deformation of Einstein--Rosen spacetime and derive the corresponding corrections to the metric and connection coefficients. In Sec.~\ref{sec:GFTP} we establish the perturbative regime considered throughout the paper and discuss general structural properties of test-particle dynamics. Monochromatic solutions are analyzed in Sec.~\ref{sec:TPM}, while the Weber--Wheeler pulse is studied in Sec.~\ref{sec:TPWW}. Final remarks and conclusions are presented in Sec.~\ref{sec:conc}. Throughout this work, we use natural units $G=c=1$.

\section{Riemannian geometry of noncommutative surfaces} \label{sec:RGNS}

In the approach of \cite{Chaichian:2006ht}, noncommutativity is incorporated geometrically through a deformation of the embedding structure of spacetime, rather than by modifying the Einstein--Hilbert action itself. In this section we briefly review this approach, emphasizing the ingredients required for explicit calculations. We refer the reader to \cite{Zhang:2008vk} for an extended review and for its formulation in terms of projective modules, thereby placing the present construction within the standard algebraic framework of noncommutative differential geometry \cite{Connes1994}.

Our starting point is the deformation quantization formalism \cite{Kontsevich:1997vb} applied to the algebra of functions on spacetime, restricting ourselves to constant Poisson structures. More precisely, let $(M,g)$ be an $n$-dimensional pseudo-Riemannian manifold describing a gravitational background, let $\{x^i\}$ be a local coordinate system defined on an open set $U\subset M$, and let $h$ be a formal deformation parameter. We consider the set $\mathcal{A}$ of formal power series in $h$,
\begin{equation}
\mathcal{A}=\left\{\sum_{i\ge0}f_i h^i \;\middle|\; f_i\in C^\infty(U)\right\},
\end{equation}
which naturally carries the structure of an $\mathbb{R}[[h]]$-module. The noncommutative structure is introduced through the Moyal star product,
\begin{equation}
u* v
=\lim_{x'\to x}
\exp\!\left(h\theta^{ij}\partial_i\partial_j'\right)u(x)v(x'),
\end{equation}
where $\theta^{ij}$ are the components of a constant antisymmetric matrix representing the Poisson tensor. The algebra $\mathcal{A}$ equipped with this product becomes associative and noncommutative, reproducing the commutation relations
\begin{equation}
[x_i,x_j]=x_i*x_j-x_j*x_i=2h\theta^{ij}.
\end{equation}
For suitable choices of $h$ and $\theta^{ij}$, these relations imply the spacetime uncertainty relations discussed in \cite{Doplicher1995}. At this stage, only the differentiable structure of $M$ is deformed.

In order to construct a corresponding deformation of the geometric structure, we follow the approach developed in \cite{Chaichian:2006ht}. One considers an isometric embedding
\begin{equation}
X:U\rightarrow \mathbb{R}^{p,q},
\end{equation}
where $\mathbb{R}^{p,q}$ denotes the pseudo-Euclidean space endowed with the scalar product 
\begin{equation}
a\cdot b=-\sum_{k=1}^{p}a^k b^k+\sum_{k=p+1}^{q+p}a^k b^k,
\end{equation}
for vectors $a=a^k e_k$ and $b=b^k e_k$. The existence of such an embedding is guaranteed by the pseudo-Riemannian generalization of Nash's embedding theorem. In terms of the embedding coordinates, the metric takes the standard form
\begin{equation}
g_{\mu\nu}=\frac{\partial X}{\partial x^\mu}\cdot\frac{\partial X}{\partial x^\nu}.
\end{equation}

The next step is to promote the embedding geometry to the noncommutative algebra $\mathcal{A}$. Consider the $\mathcal{A}$-bimodule $\mathcal{A}^m$, with $m=p+q$, together with the deformed inner product
\begin{equation}
\bullet:\mathcal{A}^m\otimes_{\mathbb{R}[[h]]}\mathcal{A}^m\rightarrow\mathcal{A},
\end{equation}
defined by
\begin{equation}
A\bullet B=-\sum_{k=1}^{p}a^k*b^k+\sum_{k=p+1}^{q+p}a^k*b^k,
\end{equation}
where $A=(a_1,\dots,a_m)$ and $B=(b_1,\dots,b_m)$ belong to $\mathcal{A}^m$. Writing
\begin{equation}
E_\mu=\frac{\partial X}{\partial x^\mu},
\end{equation}
the noncommutative metric is defined as
\begin{equation}\label{metrica}
\hat{g}_{\mu\nu}=E_\mu\bullet E_\nu.
\end{equation}
From this point onward, we denote the noncommutative metric simply by $g_{\mu\nu}$.

Let $g=(g_{\mu\nu})$ be the corresponding $n\times n$ matrix with entries in $\mathcal{A}$. Since $g\bmod h$ coincides with the classical metric and is therefore invertible, one has $g\in GL_n(\mathcal{A})$. Associativity of $\mathcal{A}$ then guarantees the existence of a unique inverse matrix $g^{\mu\nu}$ satisfying 
\begin{equation}
g_{\mu\nu}*g^{\nu\sigma}
=
g^{\sigma\nu}*g_{\nu\mu}
=
\delta_\mu^\sigma.
\end{equation}
Hence, despite the noncommutativity of the algebra, left and right inverses coincide. An explicit recursive construction for $g^{\mu\nu}$ order by order in $h$ was derived in \cite{Chaichian:2006ht}.

Following \cite{Chaichian:2006ht}, one may define a left connection $\nabla_\mu$ on the left tangent bundle
\begin{equation}
TX=\{a^\mu*E_\mu\;|\; a^\mu\in\mathcal{A}\},
\end{equation}
by composing the ordinary partial derivative with the projection onto $TX$. Explicitly,
\begin{equation}
\nabla_\mu E_\nu
=
\Gamma_{\mu\nu}^{\sigma}*E_\sigma,
\end{equation}
where
\begin{equation}\label{coefconec}
\Gamma_{\mu\nu}^{\sigma}
=
\partial_\mu E_\nu\bullet \bar{E}^\sigma,
\end{equation}
with
\begin{equation}
\bar{E}^\sigma=E_\kappa*g^{\kappa\sigma}.
\end{equation}
An analogous right connection may be defined on the right tangent bundle
\begin{equation}
\bar{T}X=\{E_\mu*a^\mu\;|\; a^\mu\in\mathcal{A}\}.
\end{equation}
The corresponding connection coefficients are symmetric in the lower indices and satisfy a noncommutative analogue of metric compatibility. In general, the left and right connections do not coincide, although both reduce to the Levi-Civita connection in the commutative limit $h\to0$. 

The Riemann tensor associated with the connection $\Gamma_{\mu\nu}^{\sigma}$ is given by
\begin{equation}
R^{\alpha}{}_{\beta\gamma\lambda}
=
\partial_\gamma\Gamma^{\alpha}{}_{\lambda\beta}
-
\partial_\lambda\Gamma^{\alpha}{}_{\gamma\beta}
+
\Gamma^{\sigma}{}_{\lambda\beta}*\Gamma^{\alpha}{}_{\gamma\sigma}
-
\Gamma^{\sigma}{}_{\gamma\beta}*\Gamma^{\alpha}{}_{\lambda\sigma}.
\end{equation}
There is a corresponding Riemann tensor on the right tangent bundle. The noncommutative geometries associated with the left and right modules are 
usually regarded as equivalent in this formalism, and the corresponding 
Riemann tensors coincide \cite{Chaichian:2006ht,Wang:2009ju,SotoCampos2018}. 
Accordingly, following the convention adopted in previous applications, we 
work with the left tangent module and its associated connection throughout this paper. Defining
\begin{equation}
R_{\kappa\beta\gamma\lambda}
=
R_{\beta\gamma\lambda}^{\alpha}*g_{\alpha\kappa},
\end{equation}
one finds
\begin{equation}
R_{\kappa\beta\gamma\lambda}
=
-
R_{\beta\kappa\lambda\gamma},
\end{equation}
although, unlike the commutative case, no simple symmetry relates $R_{\kappa\beta\gamma\lambda}$ and $R_{\beta\kappa\gamma\lambda}$.

An important consequence of noncommutativity is that the two standard contractions of the Riemann tensor no longer coincide. In the commutative case one has
\begin{equation}
R^{\alpha}{}_{\beta}
=
g^{\alpha\kappa}R^{\sigma}{}_{\kappa\sigma\beta}
=
g^{\sigma\kappa}R^{\alpha}{}_{\kappa\beta\sigma},
\end{equation}
whereas in the noncommutative setting the two contractions define distinct objects. One therefore introduces
\begin{equation}
R^{\alpha}{}_{\beta}
=
g^{\alpha\kappa}* R^{\sigma}{}_{\kappa\sigma\beta},
\end{equation}
and
\begin{equation}
\Theta^{\alpha}{}_{\beta}
=
g^{\sigma\kappa}* R^{\alpha}{}_{\kappa\beta\sigma}.
\end{equation}
Their traces nevertheless coincide,
\begin{equation}
R^{\alpha}{}_{\alpha}
=
\Theta^{\alpha}{}_{\alpha}
=
R,
\end{equation}
thus defining a unique scalar curvature.

Finally, we note that \cite{Chaichian:2006ht} also developed a noncommutative analogue of the second Bianchi identity, from which an object playing the role of the Einstein tensor can be constructed. This provides the motivation for proposing a noncommutative version of the vacuum Einstein equations within this framework.

\section{Noncommutative deformation of Einstein--Rosen geometry} \label{sec:NCERW}

In this section, we construct the noncommutative deformation of the Einstein--Rosen spacetime within the geometric framework reviewed in Sec.~\ref{sec:RGNS}. We first introduce the classical Einstein--Rosen solution and its isometric embedding, and then specify the noncommutative structure and derive the resulting metric and affine connection. These quantities determine the test-particle dynamics studied in the following sections.

The present work follows the strategy adopted in several previous gravitational applications of the formalism \cite{Wang:2008ut,SotoCampos2018,Sun:2010nas}, in which the geometry generated by the embedding construction is taken as the starting point for studying physical effects. Accordingly, our analysis relies only on the resulting metric and connection, without claiming that the deformed Einstein--Rosen geometry solves a specified set of noncommutative gravitational field equations. This distinction is important because, although the formalism admits a noncommutative analogue of the second Bianchi identity and an Einstein-tensor-like object can be constructed \cite{Chaichian:2006ht}, different choices of field equations and matter-sector prescriptions may lead to inequivalent dynamical completions. 

Einstein--Rosen waves (ERW) constitute the first explicit demonstration that General Relativity admits gravitational-wave solutions \cite{Griffiths:2009dfa,Einstein:1937qu}. They describe exact vacuum spacetimes with cylindrical symmetry. Throughout this section, we denote by $x^\mu=(x^0,x^1,x^2,x^3)=(t,\rho, \phi,z)$ the dimensionless coordinates entering the isometric embedding. Their relation with the conventional dimensionful cylindrical coordinates will be discussed below. With this notation, ERW are determined by the metric 
\begin{equation}
    ds^2=e^{2 \gamma - 2 \psi}(dt^2-d \rho^2)-\rho^2 e^{-2 \psi}d \phi^2-e^{2 \psi}dz^2,
\end{equation}
where $\gamma=\gamma(t,\rho)$ and $\psi=\psi(t,\rho)$ satisfy
\begin{equation}
    \psi_{tt}-\frac{1}{\rho} \left( \rho \psi_\rho \right)_\rho=0,
\end{equation}
together with
\begin{equation}
    \gamma_t=2 \rho \psi_t \psi_\rho,
\end{equation}
and
\begin{equation}
    \gamma_\rho= \rho \left( \psi_\rho^2 + \psi_t^2 \right).
\end{equation}

In order to implement the noncommutative deformation described in the previous section, we make use of the explicit isometric embedding constructed in \cite{ROSEN1965}. The Einstein--Rosen spacetime can be embedded into the pseudo-Euclidean space $\mathbb{R}^{4,6}$ endowed with the metric $\eta_{\alpha\beta}=\mathrm{diag}(1,1,1,1,-1,-1,-1,-1,-1,-1)$ through the map $X:\mathbb{R}^4\to\mathbb{R}^{4,6}$ given by
\begin{equation}
    X(t,\rho,\phi,z)=
    \begin{bmatrix}
        \rho \exp(- \psi) \\
        \exp(\psi) \\
        \exp(\gamma - \psi) \cos t \\
        \exp(\gamma - \psi) \sin t \\
        \exp(\gamma - \psi) \cos \rho \\
        \exp(\gamma - \psi) \sin \rho \\
        \rho \exp(- \psi) \cos \phi \\
        \rho \exp(- \psi) \sin \phi \\
        \exp (\psi) \cos z \\
        \exp(\psi) \sin z
    \end{bmatrix}.
\end{equation}

A crucial aspect of the construction is that the noncommutative deformation acts directly at the level of the embedding geometry. All geometric quantities are obtained from the embedded coordinates through the replacement of the ordinary inner product by its noncommutative counterpart. Consequently, dimensional consistency of the embedding map becomes essential. Since the coordinates $(t,\rho,\phi,z)$ appear as arguments of trigonometric functions, they must be dimensionless variables. The same applies to the arguments entering the functions $\psi$ and $\gamma$. Therefore, the coordinates appearing in the embedding should be understood as dimensionless variables obtained from the conventional cylindrical coordinates through appropriate characteristic length scales associated with the gravitational configuration under consideration. In practice, these scales arise naturally from integration constants of the Einstein equations.

Introducing conventional dimensionful cylindrical coordinates $(\tilde{t},\tilde{\rho},\tilde{\phi},\tilde{z})$, a particularly relevant example is provided by the monochromatic Einstein--Rosen wave \cite{Stephani2003}, obtained by separation of variables in the equation for $\psi$,
\begin{equation}
    \psi(\tilde{t},\tilde{\rho})=\varepsilon J_0(k \tilde{\rho}) \cos(k \tilde{t}),
\end{equation}
where $J_0$ denotes the Bessel function of the first kind of order zero, $k$ has dimensions of inverse length and $\varepsilon$ is the dimensionless wave amplitude. The quantity $k^{-1}$ therefore defines the characteristic scale of the system. In this way, a dimensionless coordinate system is obtained through 
\begin{equation}
    t=k\tilde{t}, \qquad
    \rho=k\tilde{\rho}, \qquad
    \phi=\tilde{\phi}, \qquad
    z=k\tilde{z}.
\end{equation}

Since both $\psi$ and $\gamma$ appear inside exponential functions, they are dimensionless quantities, implying that $\varepsilon$ is also dimensionless. In these variables the solution takes the form
\begin{equation}
    \psi(t,\rho)=\varepsilon J_0(\rho)\cos(t).
\end{equation}

To construct the noncommutative Einstein--Rosen geometry, it remains to specify the deformation of the algebra of functions. Here we focus on space-space noncommutativity only. Since the coordinates introduced above are dimensionless, it is natural to consider a dimensionless Poisson tensor.

The most general space-space Poisson bivector invariant under the action of the Killing fields of the Einstein--Rosen solution, $\partial_z$ and $\partial_\phi$, is
\begin{equation}
    \Theta
    =
    A(t,\rho)\,\partial_\rho\wedge\partial_z
    +
    B(t,\rho)\,\partial_\rho\wedge\partial_\phi
    +
    C(t,\rho)\,\partial_\phi\wedge\partial_z,
\end{equation}
where the coefficient functions are constrained only by the Jacobi identity. In addition to these continuous symmetries, the Einstein--Rosen spacetime is invariant under the discrete reflections $z\to -z$ and $\phi\to -\phi$. Requiring the Poisson structure to be invariant under each reflection separately, however, leaves only the trivial possibility $\Theta=0$. We adopt the angular reflection symmetry $\phi\to -\phi$ as the preferred discrete symmetry. This reduces the Poisson bivector to
\begin{equation}
    \Theta
    =
    A(t,\rho)\,\partial_\rho\wedge\partial_z.
\end{equation}

This symmetry-guided choice is in line with related constructions in which the noncommutative or Poisson structure is constrained by the symmetries of the underlying commutative geometry; see, for example, \cite{Buric:2008th,Dolan:2006hv,schupp2009}. Within this class, the simplest nontrivial choice is obtained by taking $A(t,\rho)$ to be constant, namely $\theta^{13}=-\theta^{31}=1$ with all remaining components vanishing. The corresponding Poisson tensor $\theta^{\alpha\beta}$ is represented by the antisymmetric matrix
\begin{equation}
\theta^{\alpha\beta}
=
    \begin{bmatrix}
        0 & 0 & 0 & 0 \\
        0 & 0 & 0 & 1 \\
        0 & 0 & 0 & 0 \\
        0 & -1 & 0 & 0
    \end{bmatrix}.
\end{equation}

The corresponding Moyal product is then
\begin{equation}
    f(x)*g(x)= \lim_{x' \to x} e^{h(\partial_\rho \partial_{z'} - \partial_z \partial_{\rho'})}f(x)g(x'),
\end{equation}
which reproduces the commutation relations
\begin{equation}
	[\rho,z]=2h, \qquad	[t,\rho]=[t,\phi]=[t,z]=[\rho,\phi]=[\phi,z]=0.
\end{equation}

Accordingly, h is also dimensionless, and the commutation relations imply the uncertainty relation
\begin{equation}\label{uncer.relat}
    \Delta \rho \Delta z \ge h .
\end{equation}

In terms of the original dimensionful cylindrical coordinates, one has
\begin{equation}
    [\tilde{\rho},\tilde{z}]=2\tilde{h},
\end{equation}
where $\tilde{h}$ has dimensions of length squared. For the monochromatic solution considered above, consistency with the dimensionless variables requires
\begin{equation}
    h=\tilde{h}k^2.
\end{equation}

As a consequence, all noncommutative effects depend on the dimensionless combination $\tilde{h}k^2$, making explicit how the characteristic wavelength of the gravitational configuration controls the strength of the deformation. If the dimensional deformation scale $\tilde{h}$ is associated with a microscopic length squared, this dimensionless parameter is strongly suppressed for macroscopic wavelengths. In the present work we therefore treat $h$ as a formal deformation parameter and focus on the structural dependence of the induced dynamics on the characteristic scale of the gravitational configuration.

The noncommutative metric is now obtained by replacing the ordinary inner product in the embedding construction by the deformed product $\bullet$. Although the calculation can be performed exactly, throughout this work we restrict the analysis to first order in the noncommutative parameter $h$. In this regime the star product reduces to
\begin{equation}
    f*g=fg+h \left( f_\rho g_z-f_z g_\rho \right)+\mathcal{O}(h^2).
\end{equation}

A straightforward, although lengthy, computation using Eq.~\eqref{metrica} yields the deformed metric 
\begin{equation}
    \begin{bmatrix}
        e^{2\gamma-2\psi} & 0 & 0 & he^{2\psi}\left( \psi_{t \rho}
        +2 \psi_t \psi_\rho \right) \\
        0 & -e^{2\gamma-2\psi} & 0 & he^{2\psi}\left( \psi_{\rho \rho}+2 \psi_\rho^2 \right) \\
        0 & 0 & -\rho^2e^{-2\psi} & 0 \\
        -he^{2\psi}\left( \psi_{t \rho}+2 \psi_t \psi_\rho \right) &
        -he^{2\psi}\left( \psi_{\rho \rho}+2 \psi_\rho^2 \right) & 0 & -e^{2\psi}
    \end{bmatrix}.
\end{equation}

Since the metric coefficients are independent of $z$, the star product reduces to the ordinary product in this sector. Consequently, the noncommutative inverse metric $g^{\mu\nu}$ can be computed via a standard matrix inversion up to first order $\mathcal{O}(h)$, yielding nonvanishing off-diagonal components only in the $t-z$ and $\rho-z$ sectors, which explicitly feed the deformed connection coefficients presented below. The explicit expression is omitted since it is not used below.

The deformed connection coefficients are then obtained from Eq.~\eqref{coefconec}. At first order in $h$, the noncommutative corrections appear exclusively in components involving the axial direction $z$. In particular, the deformation generates new couplings between the radial and longitudinal sectors, while the purely $(t,\rho)$ dynamics remains unchanged at this order. The induced coefficients are
\begin{equation}\label{coefconER}
\begin{alignedat}{2}
    \Gamma^{0}{}_{03}
    &=-h\exp(-2\gamma+4\psi)\psi_t\psi_{t\rho},
    &\hspace{4em}
    \Gamma^{0}{}_{13}
    &=-h\exp(-2\gamma+4\psi)\psi_t\psi_{\rho\rho},
    \\
    \Gamma^{1}{}_{03}
    &=h\exp(-2\gamma+4\psi)\psi_\rho\psi_{t\rho},
    &
    \Gamma^{1}{}_{13}
    &=h\exp(-2\gamma+4\psi)\psi_\rho\psi_{\rho\rho},
    \\
    \Gamma^{3}{}_{00}
    &=hF_1,
    &
    \Gamma^{3}{}_{01}
    &=hF_2,
    \\
    \Gamma^{3}{}_{11}
    &=hF_3,
    &
    \Gamma^{3}{}_{22}
    &=-h\rho\exp(-2\gamma)F_4,
    \\
    \Gamma^{3}{}_{33}
    &=2h\psi_\rho-h\exp(-2\gamma+4\psi)F_5.
    &&
\end{alignedat}
\end{equation}
where the functions $F_i$ denote algebraic combinations of derivatives of $\psi$ and $\gamma$. Their explicit expressions, together with the complete list of connection coefficients, are collected in Appendix~\ref{apendix}.

The structure above already exhibits the central dynamical feature of the deformation: it generates a coupling between radial gravitational propagation and the longitudinal direction, while leaving the angular sector unaffected within the present approximation.

Finally, we emphasize that the connection determined by these coefficients does not, in general, coincide with the Levi--Civita connection associated with the deformed metric. This is expected, since the existence and uniqueness of the Levi--Civita connection rely on the symmetry of the metric tensor under index permutation ($g_{\mu\nu} = g_{\nu\mu}$), which is no longer guaranteed in the noncommutative setting.

\section{General features of nonrelativistic test particles}\label{sec:GFTP}

Having constructed the noncommutative geometry, we now investigate its phenomenological consequences through the motion of test particles. Our goal is to identify the leading dynamical effects induced by the deformation and to establish the nonrelativistic approximation that will be employed throughout the remainder of the paper.

In order to extract phenomenological consequences of the noncommutative deformation derived in the previous section, we investigate the motion of test particles assuming that their trajectories are described by autoparallel curves associated with the left connection introduced in Sec.~\ref{sec:RGNS}. Although no universally accepted prescription for test-particle motion exists within this noncommutative geometric framework, autoparallels provide the direct covariant generalization of the principle of inertia, namely vanishing covariant acceleration in the absence of non-gravitational forces. This choice has also been adopted or explicitly investigated in several non-Riemannian extensions of General Relativity, including nonsymmetric gravitational theory \cite{Moffat1979,Lgar1996}, Riemann--Cartan geometries and the nonholonomic mapping principle \cite{Kleinert1999,Kleinert2000}, projective approaches to free fall \cite{Ehlers2012}, and symmetric teleparallel/nonmetricity-based models \cite{Pala2022}. In the present setting, this prescription is further consistent with the left-module convention adopted in previous applications of the formalism and reduces to the standard geodesic equation in the commutative limit.

Moreover, for the specific deformation considered here, the use of autoparallels is particularly well suited to capture the leading phenomenological effects. To first order in the noncommutative parameter, the metric acquires only antisymmetric corrections, which do not contribute to the line element $ds^2=g_{\mu\nu}dx^\mu dx^\nu$, since the ordinary product $dx^\mu dx^\nu$ is symmetric under the interchange of indices. Consequently, metric extremals obtained from the arc-length functional remain unchanged at this order. In contrast, the connection receives nontrivial first-order corrections, leading to genuinely modified autoparallel curves. We therefore use autoparallels phenomenologically, as a probe of the affine-connection structure induced by the noncommutative deformation, in the standard sense of affine geometry \cite{eisenhart1927non}, without implying that this prescription is unique in arbitrary metric-affine theories.

The particle trajectories are thus determined by the autoparallel equation
\begin{equation}
\nabla_{\dot{\gamma}}\dot{\gamma}=0,
\end{equation}
where $\gamma(\lambda)$ denotes the particle worldline, $\dot{\gamma}$ its tangent vector, and $\lambda$ an affine parameter along the curve. In local coordinates this equation takes the form
\begin{equation}\label{eq:parallel}
\ddot{x}^{\mu}+\Gamma_{\alpha \beta}^{\mu}\dot{x}^{\alpha} \dot{x}^{\beta}=0,
\end{equation}
where dots denote differentiation with respect to $\lambda$.

To obtain analytically tractable expressions with a transparent physical interpretation, we restrict our analysis to the nonrelativistic and weak-field regime. We choose the coordinate time $t=x^0$ as the evolution parameter. Assuming $\dot t\neq0$, taking the spatial components, $\mu=i$, in Eq.~\eqref{eq:parallel}, we find
\begin{equation}
    \frac{d^2x^i}{dt^2}
    +\frac{dx^i}{dt}\frac{\ddot t}{\dot t^2}
    +\Gamma^i_{00}
    +2\Gamma^i_{0j}\frac{dx^j}{dt}
    +\Gamma^i_{jk}\frac{dx^j}{dt}\frac{dx^k}{dt}
    =0.
\end{equation}

Taking instead the temporal component, $\mu=0$, in  Eq.~\eqref{eq:parallel}, we obtain
\begin{equation}
    \frac{\ddot t}{\dot t^2}
    =
    -\Gamma^0_{00}
    -2\Gamma^0_{0j}\frac{dx^j}{dt}
    -\Gamma^0_{jk}\frac{dx^j}{dt}\frac{dx^k}{dt},
\end{equation}
so that the second term in the previous equation is itself proportional to the spatial velocities. Therefore, we have
\begin{equation}
    \frac{d^2x^i}{dt^2}
    +\Gamma^i_{00}
    + \left(2\Gamma^i_{0j}\frac{dx^j}{dt} -\Gamma^0_{00}\frac{dx^i}{dt}\right)
    +\left(\Gamma^i_{jk}\frac{dx^j}{dt}\frac{dx^k}{dt}-2\Gamma^0_{0j}\frac{dx^j}{dt}\frac{dx^i}{dt} \right)
    -\Gamma^0_{jk}\frac{dx^j}{dt}\frac{dx^k}{dt}\frac{dx^i}{dt} 
    =0 \,.
\end{equation}

In the leading nonrelativistic approximation to the equations of motion, we retain the velocity-independent connection terms $\Gamma^i_{00}$ together with the purely kinematical cylindrical terms required to describe inertial motion in curvilinear coordinates. The discarded terms are linear, quadratic or cubic in the spatial velocities. In the commutative limit, the linear contributions reduce, in the linearized approximation with a time-independent Newtonian potential, to the gravitomagnetic contributions familiar from General Relativity, which constitute the leading velocity-dependent relativistic corrections to the nonrelativistic approximation \cite{ryder2009introduction,hobson2006general}. More generally, they belong to a broader class of relativistic velocity-dependent effects. The quadratic and cubic terms constitute higher-order relativistic corrections. The resulting equations are
\begin{equation}\label{din.nonrel}
    \begin{aligned}
        \rho''-\rho(\phi')^2+\Gamma^{1}{}_{00} &= 0,\\
        \phi''+\frac{2}{\rho}\rho'\phi'+\Gamma^{2}{}_{00} &= 0,\\
        z''+\Gamma^{3}{}_{00} &= 0,
    \end{aligned}
\end{equation}
where we denote derivatives with respect to $t$ by primes. The weak-field approximation requires a more careful discussion. For Einstein--Rosen waves, the perturbative hierarchy follows directly from the structure of the Einstein equations. Introducing a small dimensionless amplitude parameter $\varepsilon$, one has
\begin{equation}
    \psi = \mathcal{O}(\varepsilon),
    \qquad
    \gamma = \mathcal{O}(\varepsilon^2),
\end{equation}
since the field equation for $\psi$ is linear, whereas $\gamma$ is sourced quadratically by derivatives of $\psi$. Consequently, the metric admits an expansion around Minkowski spacetime schematically of the form
\begin{equation}
    g_{\mu\nu}
    =
    \eta_{\mu\nu}
    +
    \mathcal{O}(\psi)
    +
    \mathcal{O}(\gamma,\psi^2).
\end{equation}

Although the noncommutative deformation was formally constructed perturbatively in the parameter $h$, we find that noncommutative contributions arise only through mixed terms proportional to $h\varepsilon$, with no pure $\mathcal{O}(h)$ corrections. Therefore, consistently retaining the leading noncommutative effects requires including purely gravitational contributions at the same perturbative order, namely the $\mathcal{O}(\varepsilon^2)$ terms associated with $\gamma$ and $\psi^2$.

From a physical perspective, the $\mathcal{O}(\varepsilon^2)$ sector encodes nonlinear gravitational effects of the Einstein--Rosen wave, while the $\mathcal{O}(h\varepsilon)$ terms represent the leading noncommutative corrections triggered by the gravitational-wave background itself. In this sense, noncommutativity does not contribute as an independent sector within the present model, but instead couples effectively to the gravitational degrees of freedom responsible for the wave dynamics.

Throughout the analysis, the commutative sector of the geometry is retained through $\mathcal{O}(\varepsilon^2)$, whereas the noncommutative sector is restricted to its leading nontrivial contribution, $\mathcal{O}(h\varepsilon)$. The resulting autoparallel equations are subsequently treated as a reduced dynamical system and are solved without imposing a regular truncation of the trajectories at the same nominal order. Consequently, nonlinear iterations of the retained terms may generate observables of higher combined order. In particular, the $\mathcal{O}(h\varepsilon^2)$ longitudinal drift obtained below arises from the leading $\mathcal{O}(h\varepsilon)$ noncommutative force evaluated along the $\mathcal{O}(\varepsilon)$ classical radial response. It should therefore be interpreted as the iterated contribution predicted by the reduced dynamics, rather than as the complete coefficient of the full geometry at $\mathcal{O}(h\varepsilon^2)$. Direct connection terms at that order could, in principle, modify its numerical coefficient. This truncation is adopted deliberately in order to isolate the leading radial--longitudinal coupling and to retain both analytical tractability and a transparent interpretation of the underlying mechanism.

So, at order $\mathcal{O}(h\varepsilon)$, the equations governing the nonrelativistic dynamics in the weak-field approximation become 
\begin{equation}
    \begin{aligned}
        \rho''-\rho(\phi')^2+\gamma_\rho-\psi_\rho &= 0,\\
        (\rho^2\phi')' &= 0,\\
        z''-h\psi_{tt\rho} &= 0.
    \end{aligned}
\end{equation}

The third equation shows that the noncommutative deformation induces an effective acceleration along the axial direction absent in the commutative theory, while the radial dynamics remains unchanged at this perturbative order. In the commutative Einstein--Rosen spacetime, the Killing vector $\partial_z$ gives the conserved covariant momentum
\begin{equation}
    P_z=g_{zz} \dot{z}=-e^{2\psi}\dot{z}.
\end{equation}

The nonrelativistic limit with $\Gamma_{00}^3=0$ immediately gives $z''=0$, implying conservation of the nonrelativistic axial momentum $P^{NR}_z=z'$. The noncommutative correction modifies this structure by introducing an effective force term in the axial direction, and a direct calculation shows that $P_z$ ceases to be a first integral of the autoparallel system. The effect therefore appears as a genuine deformation of the relativistic dynamics whose phenomenological content becomes particularly transparent in the nonrelativistic regime.

Since the angular sector is unaffected by the noncommutative corrections at this order, its first integral is
\begin{equation}
    \rho^2\phi'=L.
\end{equation}
We restrict attention to zero-angular-momentum trajectories, $L=0$. As long as
$\rho>0$, axial symmetry then allows the constant value of $\phi$ to be set
to zero. The equations of motion reduce to
\begin{equation}\label{dingeral}
    \begin{aligned}
        \rho''+\gamma_\rho-\psi_\rho &= 0,\\
        z''-h\psi_{tt\rho} &= 0.
    \end{aligned}
\end{equation}

The nonrelativistic limit therefore isolates the dominant dynamical effects relevant for phenomenology while preserving the essential structure of the noncommutative correction. In particular, the central mechanism already emerges through the second time derivative of $\psi$ and does not rely on higher-order corrections. The induced longitudinal motion and its spectral dependence thus appear as robust features of the model rather than artifacts of the perturbative truncation.

To make this structure more explicit, consider a generic Einstein--Rosen solution represented as a superposition of monochromatic modes,
\begin{equation}
    \psi(t,\rho)= \varepsilon\int A(k)J_0(k \rho) \cos(kt)\,dk,
\end{equation}
where $A(k)$ specifies the spectral profile of the wave.

This representation allows the terms appearing in the equations of motion to be written directly in terms of the spectral decomposition. The radial dynamics receives contributions from both $\gamma_\rho$ and $\psi_\rho$. Since $\gamma$ is quadratic in $A(k)$, it involves mode mixing and cannot be expressed as a simple spectral functional of $A(k)$. By contrast, both $\psi_\rho$ and the noncommutative correction in the axial equation are generated by linear operators acting on $A(k)$, allowing for a direct spectral interpretation. Although the $\gamma$ contribution enters at the same perturbative order, its nonlinear structure does not affect the mechanism induced by the noncommutative sector, which remains controlled by a linear spectral weighting.

The longitudinal response therefore probes a higher spectral moment of the gravitational-wave profile than the radial motion. More precisely, while the radial dynamics receives both linear and nonlinear contributions leading to mode mixing, the noncommutative correction to the axial motion is governed directly by a well-defined linear spectral weighting. The second time derivative introduces an additional factor of $k^2$, making the longitudinal response significantly more sensitive to the high-frequency components of the wave profile. In this sense, the noncommutative sector introduces an effective observable that acts as a UV-sensitive spectral selector. This structure becomes particularly transparent in monochromatic configurations, where the induced dynamics can be analyzed analytically.

\subsection{Monochromatic waves} \label{sec:TPM}

We first consider monochromatic Einstein--Rosen waves, which provide the simplest setting in which the dynamical consequences of the noncommutative deformation can be analyzed analytically. This case makes the coupling between radial and longitudinal motion particularly transparent and illustrates the ultraviolet sensitivity discussed in the previous section.

A monochromatic Einstein--Rosen wave corresponds to the trivial spectral distribution
\begin{equation}
    A(k)=\delta(k-k_0).
\end{equation}
As discussed previously, the solution must be expressed in dimensionless variables scaled by the characteristic length $k^{-1}$. The metric function $\psi$ is therefore written as
\begin{equation}
    \psi(t,\rho)=\varepsilon J_0(\rho) \cos(t),
\end{equation}
which implies
\begin{equation}
    \gamma(t,\rho)=-\frac{1}{2}\varepsilon^2 \rho J_0( \rho) J_1( \rho) \cos(2t) + \frac{1}{4} \varepsilon^2 \rho^2 \left( J_0^2( \rho) - J_2( \rho) J_0( \rho) +2J_1^2( \rho) \right). 
\end{equation}

Substituting these expressions into Eq.~\eqref{dingeral}, we obtain
\begin{equation}
    \begin{aligned}
        \rho''
        +\varepsilon^2\rho
        \left[\cos^2(t)J_1^2(\rho)+\sin^2(t)J_0^2(\rho)\right]
        +\varepsilon J_1(\rho)\cos(t) &= 0,\\
        z''-h\varepsilon J_1(\rho)\cos(t) &= 0.
    \end{aligned}
\end{equation}

We begin by analyzing the small-radius regime $\rho\ll1$. Expanding the Bessel functions and neglecting terms of order $\mathcal{O}(\rho^2)$, the system reduces to
\begin{equation}
    \begin{aligned}
        \rho''
        +\frac{\varepsilon^2}{2}\bigl[1-\cos(2t)\bigr]\rho
        +\frac{\varepsilon}{2}\cos(t)\rho &= 0,\\
        z''-\frac{h\varepsilon}{2}\cos(t)\rho &= 0.
    \end{aligned}
\end{equation}

Despite the successive simplifications, no closed-form analytic solution appears to be available for the resulting system. In particular, the radial equation, which remains unaffected by the noncommutative deformation at this order, corresponds to a modified Mathieu-type equation \cite{Olver:2010ouy}. Since we are considering the weak-amplitude regime $\varepsilon\ll1$, we first seek a perturbative solution of the form
\begin{equation}
    \rho=\rho_0 + \varepsilon \rho_1 + \varepsilon^2 \rho_2.
\end{equation}

Because $\rho$ appears in the axial equation multiplied by the factor $h\varepsilon$, only the leading-order contribution $\rho_0$ is required in the equation for $z$. Imposing the initial conditions
\begin{equation}
    \rho(0)=\rho_0,
    \qquad
    \rho'(0)=0,
    \qquad
    z(0)=z'(0)=0,
\end{equation}
we obtain
\begin{equation}
    \begin{aligned}
        \rho(t)
        &=\rho_0\Biggl(
            1+\frac{\varepsilon}{2}\bigl[\cos(t)-1\bigr]
            +\varepsilon^2\left[
                \frac{11}{32}-\frac{5t^2}{16}
                -\frac{\cos(t)}{4}-\frac{3\cos(2t)}{32}
            \right]
        \Biggr),\\
        z(t)
        &=\frac{h\varepsilon\rho_0}{2}\bigl[1-\cos(t)\bigr].
    \end{aligned}
\end{equation}

The perturbative solution for $\rho$ contains a secular, i.e. non-periodic and unbounded, contribution proportional to $t^2$. The exact solution, however, remains bounded. Such secular terms are well known in perturbation theory and signal the breakdown of the naive expansion at sufficiently large times. In order to construct a uniformly valid approximation, we employ the method of multiple scales \cite{Kevorkian1996}. Since the unperturbed system is not oscillatory and possesses no intrinsic natural frequency, the Poincar\'e--Lindstedt method \cite{Verhulst1996} cannot be applied.

Following the method of multiple scales, we introduce a fast time scale $T_0=t$ and a slow time scale $T_1=\varepsilon t$, and seek a solution of the form
\begin{equation}
    \rho = \rho_0(T_0,T_1)+\varepsilon \rho_1(T_0,T_1).
\end{equation}

Although the general philosophy of the method is standard, different implementations appear in the literature depending on how initial conditions are incorporated into the perturbative hierarchy \cite{Kevorkian1996,NewellMoloney2004}. Here we follow the approach described in \cite{NewellMoloney2004} and succinctly reviewed in \cite{Jakobsen}. In practice, one constructs the most general zeroth-order solution free of secular contributions and selects a particular first-order solution. The full dependence on the slow scale $T_1$ is then determined by eliminating the secular terms that would arise at second order. The multiple-scales solution provides a uniformly valid approximation to the truncated equations on the slow timescale $T_1=\varepsilon t$. The oscillatory envelope is generated by the leading noncommutative coupling retained in the equations, while higher-order terms are expected to produce subleading corrections to its frequency and phase together with higher harmonics on the fast time scale.

Carrying out this procedure yields 
\begin{equation}
    \rho = \frac{2\rho_0}{2+ \varepsilon} 
    \cos \left( \sqrt{\frac{5}{8}} \varepsilon t \right)
    \left(1+ \frac{\varepsilon}{2} \cos t \right).
\end{equation}

Substituting this expression into the equation for $z$ and integrating, we obtain, consistently to leading order in the noncommutative contribution $\mathcal{O}(h\varepsilon)$,
\begin{equation}
    z=\frac{1}{50}\rho_0h \left( (25\varepsilon + 10) 
    - \cos \left( \sqrt{\frac{5}{8}} \varepsilon t \right) 
    (25 \varepsilon \cos t+10) \right).
\end{equation}

The classical gravitational-wave therefore induces a quasi-periodic oscillatory motion in the radial direction, while the noncommutative correction generates a corresponding quasi-periodic longitudinal response directly coupled to the radial dynamics. The induced longitudinal motion selects a preferred orientation along the symmetry axis, determined by the sign of the noncommutative parameter $h$.

Since the solution is quasi-periodic, there is no unique oscillation period. We therefore define the oscillation amplitude as the supremum of the displacement from the mean position. The explicit solution gives
\begin{equation}
    A_\rho=\rho_0,
    \qquad
    A_z=\rho_0|h|\left(\frac{1}{5}+\frac{\varepsilon}{2}\right),
    \qquad
    \frac{A_z}{A_\rho}
    =|h|\left(\frac{1}{5}+\frac{\varepsilon}{2}\right).
\end{equation}

This phenomenological result is consistent with several general expectations. First, the noncommutative deformation implements the uncertainty relation Eq.~\eqref{uncer.relat} which prevents simultaneous sharp localization in the radial and axial directions. An induced oscillation in $z$ with amplitude proportional to the noncommutative scale may therefore be interpreted as an effective manifestation of this underlying geometric uncertainty.

Second, the effect grows with the wave amplitude $\varepsilon$, showing that the noncommutative response is amplified by the gravitational field itself. Finally, recalling that in dimensionful variables one has
\begin{equation}
    h=\tilde{h}k^2,
\end{equation}
the induced longitudinal amplitude becomes proportional to $k^2$, namely, to the square of the gravitational-wave frequency. The noncommutative observable $A_z$ is therefore explicitly UV sensitive. Since higher frequencies correspond to shorter characteristic length scales and higher energies, this behavior is consistent with the general expectation that noncommutative effects become more relevant in ultraviolet regimes.

We now briefly comment on the asymptotic regime $\rho\gg1$. Using the Hankel asymptotic expansions \cite{Olver:2010ouy}, the equations reduce to
\begin{equation}
    \begin{aligned}
        \rho''
        &+\frac{2\varepsilon^2}{\pi}
          \left[\frac{1}{2}-\cos(\rho)\sin(\rho)\cos(2t)\right]
          +\varepsilon\sqrt{\frac{2}{\pi\rho}}
          \cos\!\left(\rho-\frac{3\pi}{4}\right)\cos(t)=0,\\
        z''
        &-h\varepsilon\sqrt{\frac{2}{\pi\rho}}
          \cos\!\left(\rho-\frac{3\pi}{4}\right)\cos(t)=0.
    \end{aligned}
\end{equation}

The longitudinal force induced by the noncommutative deformation is therefore spatially suppressed at large distances. However, the radial equation contains an average contribution directed toward the symmetry axis through the term proportional to $\varepsilon^2$. Indeed,
\begin{equation}
    |\cos\rho \sin \rho|\le\frac{1}{2}
    \implies
    \frac{1}{2}-\cos(\rho)\sin(\rho)\cos(2t)\ge0,
\end{equation}
showing that the radial dynamics tends to drive the particle back toward smaller values of $\rho$. Consequently, even trajectories initially satisfying $\rho\gg1$ eventually probe regions where the asymptotic expansion ceases to be uniformly valid. A complete characterization of the global longitudinal response in this regime therefore becomes considerably more subtle and lies beyond the scope of the present work.

The monochromatic case nevertheless provides a particularly transparent realization of the UV-sensitive nature of the noncommutative longitudinal response. Since its spectrum is concentrated at a single frequency, however, it does not capture the spectral mixing effects characteristic of localized gravitational pulses, which we analyze in the next section.


\subsection{Weber--Wheeler pulse}\label{sec:TPWW}

We now turn to the Weber--Wheeler pulse, whose continuous spectrum allows us to investigate how the noncommutative response depends on the full spectral content of the gravitational-wave. Unlike the monochromatic case, this configuration describes a localized scattering process and therefore provides a natural setting to study memory-like effects and asymptotic observables.

The Weber--Wheeler pulse \cite{Weber1957}, later analyzed by Bonnor \cite{Bonnor1957}, is obtained from the spectral superposition
\begin{equation}
    \psi(t,\rho)=2C \int_0^{\infty}e^{-ak}J_0(k \rho)\cos(kt)\,dk,
\end{equation}
which admits the closed-form representation
\begin{equation}
    \psi(t, \rho)= \sqrt{2}C 
    \left(
    \frac{
    \big((a^2+\rho^2-t^2)^2+4a^2t^2\big)^{1/2}
    +a^2+\rho^2-t^2
    }{
    (a^2+\rho^2-t^2)^2+4a^2t^2
    }
    \right)^{\frac{1}{2}}.
\end{equation}

The corresponding function $\gamma(t,\rho)$ is determined by quadratures through the Einstein equations but will not be required explicitly in the analysis below. Physically, this solution describes a localized pulse of cylindrical gravitational radiation propagating toward the symmetry axis and subsequently reflected. In addition to cylindrical symmetry, the solution is also time-symmetric. Further discussion of its geometrical and physical properties may be found in \cite{Griffiths:2009dfa}.

As discussed previously, the noncommutative construction requires the use of dimensionless coordinates. Since the parameter $a$ defines the characteristic length scale of the pulse, we introduce the dimensionless variables
\begin{equation}
    t=\frac{\tilde{t}}{a},
    \qquad
    \rho=\frac{\tilde{\rho}}{a},
    \qquad
    \phi=\tilde{\phi},
    \qquad
    z=\frac{\tilde{z}}{a},
\end{equation}
where $(\tilde{t},\tilde{\rho},\tilde{\phi},\tilde{z})$ denote the conventional cylindrical coordinates. The function $\psi$ then takes the form
\begin{equation}
    \psi(t, \rho)= 
    \sqrt{2}\varepsilon
    \left(
    \frac{
    \big((1+\rho^2-t^2)^2+4t^2\big)^{1/2}
    +1+\rho^2-t^2
    }{
    (1+\rho^2-t^2)^2+4t^2
    }
    \right)^{\frac{1}{2}},
\end{equation}
with
\begin{equation}
    \varepsilon=\frac{C}{a}.
\end{equation}

Consistency with the dimensionless formulation also requires
\begin{equation}
    h=\frac{\tilde{h}}{a^2}.
\end{equation}

In contrast with the monochromatic configuration, the Weber--Wheeler pulse possesses a continuous spectral profile. In dimensionless form the spectrum is given by
\begin{equation}
    A(K)=2e^{-K},
\end{equation}
so that the induced longitudinal response depends nontrivially on the full spectral distribution. The resulting test-particle dynamics is therefore naturally interpreted as a scattering problem. As in the previous section, we analyze separately the regimes of small and large $\rho$.

For $\rho\ll1$, we expand the Bessel functions appearing in the integral representations of $\psi_\rho$, $\psi_t$, and $\psi_{\rho tt}$ and discard terms of order $\mathcal{O}(\rho^2)$. Once again, the noncommutative contribution introduces an additional ultraviolet weighting through higher-order time derivatives. The equations of motion, Eq.~\eqref{dingeral}, therefore become
\begin{equation}
    \begin{aligned}
        \rho''
        +\left[
            \varepsilon^2\frac{16t^2}{(1+t^2)^4}
            -\varepsilon\frac{6t^2-2}{(1+t^2)^3}
        \right]\rho &= 0,\\
        z''
        -h\varepsilon
        \frac{120t^4-240t^2+24}{(1+t^2)^5}\rho &= 0.
    \end{aligned}
\end{equation}

We study this system under the initial conditions
\begin{equation}
    \rho(t)\to\rho_0,
    \qquad
    \rho'(t)\to0,
    \qquad
    z(t),z'(t)\to0,
    \quad t\to-\infty.
\end{equation}

Physically, the particle is initially at rest while the pulse is infinitely far away. The pulse crosses the particle near $t_1\approx-\rho_0$ and again near $t_2\approx\rho_0$, so that the effective interaction interval is approximately $\Delta t\approx2\rho_0$. For $\rho_0\ll1$, this interaction time is sufficiently short to justify a regular perturbative expansion in $\varepsilon$. We therefore seek solutions of the form
\begin{equation}
    \rho=\rho_0+\varepsilon\rho_1+\varepsilon^2\rho_2,
    \qquad
    z=h(\varepsilon z_1+\varepsilon^2 z_2),
\end{equation}
which yields
\begin{equation}
    \begin{aligned}
        \rho(t)
        &=\rho_0\Biggl(
            1+\frac{\varepsilon}{1+t^2}
            +\varepsilon^2\left[
                -\frac{5}{4}t\arctan t
                -\frac{5\pi}{8}t
                +\frac{10t^2+2}{24(1+t^2)^2}
                -\frac{5}{4}
            \right]
        \Biggr),\\
        z(t)
        &=h\varepsilon\rho_0\Biggl(
            \frac{6t^2-2}{(1+t^2)^3}
            +\frac{\varepsilon}{40}\left[
                15\pi t+30+30t\arctan t
                -\frac{10t^6+34t^4-74t^2+94}{(1+t^2)^4}
            \right]
        \Biggr).
    \end{aligned}
\end{equation}

The dominant late-time behavior is therefore
\begin{equation}
    \begin{aligned}
        \rho(t)
        &=\rho_0-\frac{5\pi}{4}\varepsilon^2\rho_0\,t+o(t),\\
        z(t)
        &=\frac{3\pi}{4}h\varepsilon^2\rho_0\,t+o(t),
        \qquad t\to+\infty.
    \end{aligned}
\end{equation}

Thus, the passage of the pulse produces a permanent radial impulse, while the noncommutative deformation induces an additional longitudinal impulse absent in the commutative theory. This effect is analogous to gravitational memory \cite{Bieri2015}, in the sense that the passage of the gravitational-wave permanently modifies the asymptotic state of the test particle. In the present case, the particle acquires terminal velocities $v_\infty^\rho$ and $v_\infty^z$ satisfying
\begin{equation}
    v_\infty^z
    =
    -\frac{3}{5}h\,v_\infty^\rho.
\end{equation}

The noncommutative correction therefore generates a residual longitudinal drift directly associated with the algebraic structure
\begin{equation}
    [\rho,z]=2h.
\end{equation}

As in the monochromatic case, the longitudinal response is entirely induced by the radial dynamics, although the present configuration inherits the scattering character of the pulse. While monochromatic waves generate bounded oscillatory motion, the Weber--Wheeler pulse produces a genuine asymptotic drift.

In dimensional variables, the induced longitudinal drift is given by
\begin{equation}
    v_\infty^z
    =
    -\frac{3}{4}\pi
    \frac{\tilde{h}}{a^2}
    \frac{C^2}{a^2}
    \rho_0,
\end{equation}
showing explicitly that the effect becomes significant only when sufficiently small spatial scales coexist with sufficiently strong gravitational fields.

We now turn to the large-distance regime $\rho\gg1$. To ensure the validity of the asymptotic approximations, we consider the particle initially at rest at $t=0$, when the pulse is concentrated near the origin. In this configuration, the outgoing pulse crosses the particle only once, with the interaction localized near $t\approx\rho_0$.

Introducing the retarded variable
\begin{equation}
    x=t-\rho,
\end{equation}
and subsequently defining 
\begin{equation}
    u=\rho^{-1},
\end{equation}
the pulse admits the asymptotic expansion
\begin{equation}
    \psi(x,u)
    =
    \varepsilon u^{1/2}F(x)
    +
    \mathcal{O}(u^{3/2}),
\end{equation}
or equivalently
\begin{equation}
    \psi(t,\rho)
    \sim
    \frac{1}{\sqrt{\rho}}F(t-\rho),
\end{equation}
where
\begin{equation}
    F(x)
    =
    \sqrt{
    \frac{
    \sqrt{x^2+1}-x
    }{
    x^2+1
    }}.
\end{equation}

Substituting this expression into Eq.~\eqref{dingeral}, the radial equation becomes, at leading order,
\begin{equation}
    \rho''
    +
    2\varepsilon^2
    \left(F'(t-\rho)\right)^2
    +
    \varepsilon\rho^{-1/2}F'(t-\rho)
    =0,
\end{equation}
subject to
\begin{equation}
    \rho(0)=\rho_0,
    \qquad
    \rho'(0)=0.
\end{equation}

Since $F(x)$ decays rapidly as $|x|\to\infty$, its contribution may be regarded as effectively localized within an interval $(-L,L)$. Indeed, $F$ has the asymptotic behavior $F\sim 2^{-1/2}x^{-3/2}$ as $x\to\infty$ and $F \sim \sqrt{2}|x|^{-1/2}$ as $x\to-\infty$. The pulse therefore interacts with the particle only when
\begin{equation}
    x=t-\rho_0\in(-L,L),
\end{equation}
namely for
\begin{equation}
    t\in(\rho_0-L,\rho_0+L).
\end{equation}

During this interval,r one may approximate, to leading order in $\varepsilon$, $\rho\simeq\rho_0$ giving
\begin{equation}
    \Delta\rho'
    =
    -\int_{-\infty}^{\infty}
    \left(
    2\varepsilon^2(F')^2
    +
    \varepsilon\rho_0^{-1/2}F'
    \right)dt
    =
    -\varepsilon^2.
\end{equation}

To the best of our knowledge, this nonlinear contribution has not been reported previously in this form. In the original analysis of Weber \cite{Weber1957}, test-particle motion was treated only at linear order, for which the pulse produces a transient displacement but no permanent change in the asymptotic velocity. In the time-symmetric configuration, the ingoing and outgoing linear contributions cancel exactly. By contrast, once the second-order terms are retained, a nonvanishing terminal radial velocity emerges,
\begin{equation}
\Delta\rho'
\sim
-2\varepsilon^2
\int(F')^2dx.
\end{equation}

This is a purely commutative general-relativistic effect. Since asymptotically
\begin{equation}
\gamma_\rho\sim(F')^2,
\end{equation}
it can be interpreted as the cumulative scattering effect of the effective energy density carried by cylindrical gravitational-waves, namely the C-energy introduced by Thorne \cite{Thorne1965}. Unlike the linear displacement, this second-order contribution is not canceled by the reflection process. An analogous computation performed using the advanced variable $x=t+\rho$ and initial conditions at $t\to-\infty$ yields the same contribution,
\begin{equation}
\Delta\rho'=-\varepsilon^2.
\end{equation}

Hence, in the fully symmetric configuration, the total terminal radial velocity becomes
\begin{equation}
\Delta\rho'_{\mathrm{total}}
=
-2\varepsilon^2.
\end{equation}

This classical velocity memory effect provides the commutative scattering background on top of which the noncommutative longitudinal drift is generated.

Motivated by the success of the asymptotic approximation in the radial sector, one may attempt to apply the same procedure to the longitudinal equation,
\begin{equation}
    z''+h\varepsilon\rho_0^{-1/2}F'''(t-\rho_0)=0,
    \qquad
    z(0)=z'(0)=0.
\end{equation}

At leading order, this yields a vanishing contribution to the terminal longitudinal velocity. One must therefore retain at least
\begin{equation}
    \rho=\rho_0+\varepsilon\rho_1
\end{equation}
inside the longitudinal equation, obtaining
\begin{equation}
    z''
    =
    -h\varepsilon\rho_0^{-1/2}F'''(t-\rho_0)
    +
    h\varepsilon^2
    \left(
    \frac{\rho_0^{-3/2}}{2}\rho_1F'''
    +
    \rho_0^{-1/2}\rho_1F''''
    \right),
\end{equation}
where $\rho_1$ satisfies
\begin{equation}
    \rho_1''=-\rho_0^{-1/2}F'(t-\rho_0),
    \qquad
    \rho_1(0)=\rho_1'(0)=0.
\end{equation}

This leads to
\begin{equation}
    \Delta z'
    =
    -h\varepsilon^2
    \frac{1}{4}\rho_0^{-2}.
\end{equation}

However, the structure of the longitudinal equation indicates that the leading asymptotic approximation
\begin{equation}
    \psi\sim\rho^{-1/2}F(x)
\end{equation}
is insufficient to fully capture the dynamics of $\Delta z'$. Unlike the radial sector, the longitudinal equation involves higher-order time derivatives of $\psi$, making it substantially more sensitive to subleading asymptotic contributions. Indeed, the next term

\begin{equation}
    \psi
    \sim
    \varepsilon\left[\rho^{-1/2}F(x)+\rho^{-3/2}G(x)\right]
\end{equation}
already contributes at the same asymptotic order. However, the function $G(x)$ is no longer localized and behaves asymptotically as
\begin{equation}
    G(x)\sim\frac{1}{4}\sqrt{-2x},
    \qquad
    x\to-\infty.
\end{equation}

This indicates that the asymptotic expansion ceases to be uniformly valid in the longitudinal sector, suggesting that a quantitatively reliable description requires the full spectral representation of the solution.

Starting directly from Eq.~\eqref{dingeral}, we obtain
\begin{equation}
    \Delta z'
    =
    h\int_0^\infty
    \psi_{\rho tt}(s,\rho)\,ds.
\end{equation}
Expanding
\begin{equation}
    \rho=\rho_0+\varepsilon\rho_1,
\end{equation}
one finds
\begin{equation}
    \Delta z'
    =
    h\int_0^\infty
    \psi_{\rho tt}(s,\rho_0)
    +
    \varepsilon\rho_1
    \psi_{\rho\rho tt}(s,\rho_0)
    \,ds.
\end{equation}
The first term vanishes identically since $\psi_{\rho tt}=\partial_t (\psi_{\rho t})$ and the boundary terms vanish. For the second term, Eq.~\eqref{dingeral} implies
\begin{equation}
    \varepsilon \rho_1''
    =
    \psi_\rho(t,\rho_0),
\end{equation}
which yields
\begin{equation}
    \rho_1
    =
    2
    \int_0^\infty
    e^{-K}
    \left(-KJ_1(K\rho_0)\right)
    \left(
    \frac{1-\cos(Kt)}{K^2}
    \right)dK.
\end{equation}

Similarly,
\begin{equation}
    \psi_{\rho\rho tt}
    =
    -\varepsilon
    \int_0^\infty
    e^{-K}K^4
    \left(
    J_2(K\rho)-J_0(K\rho)
    \right)
    \cos(Kt)\,\mathrm{d}K.
\end{equation}

Substituting these expressions into $\Delta z'$ and using the distributional identities
\begin{eqnarray}
    \int_0^\infty\cos(Kt)\,dt
    &=&
    \pi\delta(K),
    \\
    \int_0^\infty
    \cos(Kt)\cos(K't)\,dt
    &=&
    \frac{\pi}{2}\delta(K-K'),
\end{eqnarray}
with $K,K'\ge0$, we obtain
\begin{equation}
    \Delta z'
    =
    -h\varepsilon^2\pi
    \int_0^\infty
    e^{-2K}
    K^3
    J_1(K\rho_0)
    \left(
    J_2(K\rho_0)-J_0(K\rho_0)
    \right)dK.
\end{equation}

The oscillatory structure of this integral shows that the longitudinal response results from delicate cancellations among distinct spectral components of the pulse. In particular, the factor $K^3$ introduces a strong ultraviolet weighting, making the longitudinal sector substantially more sensitive to high-frequency components than the radial dynamics.

This ultraviolet enhancement may be quantified more explicitly by comparing the effective spectral weights associated with the radial and longitudinal sectors. In the Weber--Wheeler profile, the corresponding effective spectral weights are proportional to
\begin{equation}
    K e^{-K}
    \qquad
    \text{and}
    \qquad
    K^3 e^{-K},
\end{equation}
respectively. While the first is maximized at
\begin{equation}
    K_{\rho}^{max}=1,
\end{equation}
the second reaches its maximum at
\begin{equation}
    K_z^{max}=3.
\end{equation}

Moreover, the corresponding mean frequencies satisfy
\begin{equation}
    \langle K\rangle_z
    =
    2\langle K\rangle_{\rho},
\end{equation}
showing that the longitudinal dynamics samples preferentially significantly higher-frequency components of the gravitational pulse.

We therefore conclude that the terminal longitudinal velocity is controlled by a genuinely nontrivial spectral combination involving products of Bessel functions and cubic spectral weights. Unlike the radial sector, whose asymptotic dynamics is largely governed by the localized retarded profile $F(x)$, the longitudinal response depends in a significantly more refined manner on the full spectral structure of the gravitational pulse. Although the asymptotic approximation correctly reproduces the leading behavior
\begin{equation}
    \Delta z'
    \sim
    -\frac{h\varepsilon^2}{4\rho_0^2},
\end{equation}
the extraction of subleading corrections reveals a considerably subtler structure involving nontrivial cancellations between different orders of the asymptotic Bessel expansion. This strongly suggests that the noncommutative longitudinal dynamics does not admit a simple asymptotic description beyond leading order. 

Within the equations truncated at $\mathcal{O}(\varepsilon^2)$ and $\mathcal{O}(h\varepsilon)$, the nonlinear coupling between the classical radial response and the leading noncommutative longitudinal force generates a net drift of order $h\varepsilon^2$. This term represents the iterated contribution generated by the retained dynamics; a complete determination of the coefficient at $\mathcal{O}(h\varepsilon^2)$ would additionally require, besides the above discussion, assessing direct contributions from connection coefficients at that order.

From a physical perspective, the cubic spectral factor $K^3$ shows explicitly that noncommutativity acts as an effective ultraviolet spectral filter, selectively amplifying the contribution of high-frequency modes. In dimensional variables, the dominant strength of the effect remains controlled by the dimensionless combination
\begin{equation}
    \frac{\tilde{h}}{a^2}\frac{C^2}{a^2},
\end{equation}
showing that the induced longitudinal response remains strongly suppressed outside simultaneously ultraviolet and strongly gravitational regimes.

\section{Conclusions} \label{sec:conc}

Although it is widely expected that a noncommutative formulation of General Relativity may reveal deviations associated with quantum gravitational effects, there is still no consensus regarding the fundamental geometric structure underlying such a theory \cite{Doplicher1995,Szabo:2006wx,Hossenfelder2013}. Several approaches have been proposed in the literature to deform the geometry itself \cite{Aschieri:2005yw,Chamseddine:2000si,Chamseddine:2003we,Chaichian:2006ht}, yet comparatively few works investigate the dynamics of physical probes, and even fewer explore their corresponding effective observables.

In this work, we employed the formalism developed in \cite{Chaichian:2006ht} to construct a noncommutative deformation of Einstein--Rosen cylindrical gravitational-waves with noncommutativity between the radial and axial coordinates. The deformed metric and connection coefficients were obtained explicitly to first order in the noncommutative parameter, without imposing weak-field approximations on the gravitational sector itself. We then investigated the motion of test particles assuming that their trajectories are governed by autoparallel curves associated with the left connection. Restricting the analysis to the effective low-velocity and weak-field regime, where the phenomenological structure of the noncommutative correction becomes more transparent, we found that the radial dynamics remains unaffected by the noncommutative deformation at the perturbative order considered. In contrast, the deformation induces an effective longitudinal dynamics absent in the commutative nonrelativistic limit. The relevant axial equation contains the combination $\partial_t^2\partial_\rho\psi$, which introduces an enhanced sensitivity to the high-frequency components of the gravitational-wave spectral profile.

For monochromatic solutions, this structure manifests itself through induced longitudinal oscillations whose amplitude is proportional to the commutative radial amplitude and weighted by an additional factor proportional to the square of the characteristic wave number. This behavior shows explicitly that the noncommutative response preferentially amplifies high-frequency modes, consistently with the general expectation that noncommutative effects become more relevant in ultraviolet regimes. In this sense, the longitudinal response acts effectively as a UV-sensitive observable induced by the underlying noncommutative structure.

For the Weber--Wheeler pulse, we analyzed separately the regimes $\rho_0\ll1$ and $\rho_0\gg1$. In the short-distance regime, we obtained explicit asymptotic terminal velocities showing that the noncommutative deformation induces a residual longitudinal component linearly coupled to the commutative radial impulse. Moreover, in the asymptotic regime, we found that the longitudinal response is governed by a nontrivial spectral combination involving products of Bessel functions together with cubic frequency weights. Unlike the radial sector, whose asymptotic dynamics is essentially controlled by the effective energetic content associated with the localized pulse profile, the longitudinal dynamics exhibits a significantly stronger sensitivity to the complete spectral composition of the gravitational solution, reflecting an enhanced weighting of ultraviolet modes.

An additional result emerging from the asymptotic analysis of the Weber--Wheeler pulse is a nonlinear radial velocity memory effect already present in ordinary commutative General Relativity. This effect appears at second order in the wave amplitude and is naturally associated with the effective energy carried by cylindrical gravitational-waves, or C-energy. In particular, it survives even in the time-symmetric ingoing--outgoing configuration, showing that the cumulative scattering effect generated by the gravitational-wave energy density is not canceled by the reflection process. The noncommutative deformation then produces an additional longitudinal drift coupled to this nonlinear commutative scattering dynamics.

Taken together, these results suggest that noncommutative corrections may encode information not only about the intensity of the gravitational field but also about the detailed spectral structure of the underlying gravitational radiation. From this perspective, noncommutativity acts effectively as a UV spectral filter, selectively enhancing the contribution of high-frequency components to the induced longitudinal dynamics. We also emphasize that the magnitude of the noncommutative parameter is not fixed by the embedding formalism itself. The dimensionless quantities that control the effects found here are of the form $\tilde h k^2$ for monochromatic waves or $\tilde h/a^2$ for localized pulses. If the dimensional parameter $\tilde h$ is tied to a microscopic fundamental scale, such as the Planck length squared $(\ell_P^2)$, these effects are overwhelmingly suppressed for macroscopic gravitational-wavelengths. The results obtained in this paper should therefore be interpreted as identifying the qualitative dynamical structure and spectral selectivity induced by the deformation, rather than as proposing an immediately observable gravitational-wave signature. Deriving model-dependent phenomenological bounds on $\tilde h$ lies beyond the scope of the present work.

Although Einstein--Rosen waves do not constitute realistic astrophysical models, their exact character allows one to investigate analytically nonlinear effects, effective dynamical mechanisms, and spectral structures that would be considerably more difficult to access in more general gravitational geometries. This role is particularly natural in the present embedding-based framework, where the explicit Rosen embedding provides a concrete route to implement the noncommutative deformation while maintaining control over the underlying radiative vacuum geometry. The present analysis should therefore be understood as a proof-of-principle study within this specific framework and in an exact nonlinear background: it shows that the deformation considered here gives rise to a UV-sensitive longitudinal response and to memory-like scattering effects in a setting where the relevant mechanisms can be followed analytically.

Several directions remain open for future investigation. In particular, it would be interesting to analyze the test-particle motion beyond the approximations adopted here, especially without assuming low velocities or weak fields. Although the complete equations present substantial analytical complexity, numerical methods may provide valuable information regarding regimes beyond the perturbative and nonrelativistic truncations considered in this work. It would also be natural to investigate the collective behavior of congruences of test particles and possible noncommutative generalizations of the Jacobi equation. It would also be worthwhile to investigate whether a qualitatively similar coupling between radial and longitudinal sectors emerges in other noncommutative approaches to gravity, such as those based on the Seiberg--Witten map \cite{Chamseddine:2000si,Chamseddine:2003we} or on direct star-product deformations of the Einstein--Hilbert action \cite{Aschieri:2005yw}, or whether it is a distinctive feature of the embedding-based construction adopted here. Finally, an important question is whether similar mechanisms persist in more realistic gravitational-wave geometries associated with astrophysical sources.

\section*{Acknowledgements}
 D.C.R. acknowledges \textit{Conselho Nacional de Desenvolvimento Científico e Tecnológico} (CNPq, Brazil) and \textit{Fundação de Amparo à Pesquisa e Inovação do Espírito Santo} (FAPES, Brazil) for partial support.
 E.S.C.F. is  supported by CNPq/PDJ 153723/2025-4.

\appendix

\section{Connection coefficients}\label{apendix}

The connection is symmetric in its lower indices,
$\Gamma^\mu{}_{\alpha\beta}=\Gamma^\mu{}_{\beta\alpha}$. Up to
$\mathcal{O}(h^2)$, its nonzero independent components are listed below; all
unlisted components vanish. For upper index $0$, one has
\begin{equation}
\begin{aligned}
    \Gamma^{0}{}_{00}=\Gamma^{0}{}_{11}
    &=\gamma_t-\psi_t,
    &
    \Gamma^{0}{}_{01}&=\gamma_\rho-\psi_\rho,
    \\
    \Gamma^{0}{}_{03}
    &=-h e^{-2\gamma+4\psi}\psi_t\psi_{t\rho},
    &
    \Gamma^{0}{}_{13}
    &=-h e^{-2\gamma+4\psi}\psi_t\psi_{\rho\rho},
    \\
    \Gamma^{0}{}_{22}&=-\rho^2e^{-2\gamma}\psi_t,
    &
    \Gamma^{0}{}_{33}&=e^{-2\gamma+4\psi}\psi_t.
\end{aligned}
\end{equation}
For upper index $1$,
\begin{equation}
\begin{aligned}
    \Gamma^{1}{}_{00}=\Gamma^{1}{}_{11}
    &=\gamma_\rho-\psi_\rho,
    &
    \Gamma^{1}{}_{01}&=\gamma_t-\psi_t,
    \\
    \Gamma^{1}{}_{03}
    &=h e^{-2\gamma+4\psi}\psi_\rho\psi_{t\rho},
    &
    \Gamma^{1}{}_{13}
    &=h e^{-2\gamma+4\psi}\psi_\rho\psi_{\rho\rho},
    \\
    \Gamma^{1}{}_{22}&=\rho e^{-2\gamma}(\rho\psi_\rho-1),
    &
    \Gamma^{1}{}_{33}&=-e^{-2\gamma+4\psi}\psi_\rho.
\end{aligned}
\end{equation}
The only nonzero components with upper index $2$ are
\begin{equation}
    \Gamma^{2}{}_{02}=-\psi_t,
    \qquad
    \Gamma^{2}{}_{12}=\rho^{-1}-\psi_\rho.
\end{equation}
Finally, for upper index $3$,
\begin{equation}
\begin{aligned}
    \Gamma^{3}{}_{00}&=hF_1,
    &
    \Gamma^{3}{}_{01}&=hF_2,
    &
    \Gamma^{3}{}_{03}&=\psi_t,
    \\
    \Gamma^{3}{}_{11}&=hF_3,
    &
    \Gamma^{3}{}_{13}&=\psi_\rho,
    &
    \Gamma^{3}{}_{22}&=-h\rho e^{-2\gamma}F_4,
    \\
    \Gamma^{3}{}_{33}
    &=2h\psi_\rho-h e^{-2\gamma+4\psi}F_5.
\end{aligned}
\end{equation}

The functions $F_i$ are
\begin{equation}
\begin{aligned}
F_1={}&-\psi_{tt\rho}-\psi_\rho\psi_{\rho\rho}
        -3\psi_t\psi_{t\rho}-2\psi_\rho\psi_{tt}
        +\gamma_\rho\psi_{\rho\rho}+\gamma_t\psi_{t\rho}
        \\
    &{}-2\psi_\rho^3-4\psi_t^2\psi_\rho
        +2\gamma_\rho\psi_\rho^2+2\gamma_t\psi_t\psi_\rho,
        \\[0.25em]
F_2={}&-\psi_{t\rho\rho}-4\psi_\rho\psi_{t\rho}
        -2\psi_t\psi_{\rho\rho}
        +\gamma_t\psi_{\rho\rho}+\gamma_\rho\psi_{t\rho}
        \\
    &{}-6\psi_t\psi_\rho^2+2\gamma_t\psi_\rho^2
        +2\gamma_\rho\psi_t\psi_\rho,
        \\[0.25em]
F_3={}&-\psi_{\rho\rho\rho}-\psi_t\psi_{t\rho}
        -5\psi_\rho\psi_{\rho\rho}-2\psi_\rho\psi_t^2-4\psi_\rho^3
        \\
    &{}+\gamma_\rho\psi_{\rho\rho}+\gamma_t\psi_{t\rho}
        +2\gamma_t\psi_t\psi_\rho+2\gamma_\rho\psi_\rho^2,
        \\[0.25em]
F_4={}&2\rho\psi_\rho\psi_t^2+\rho\psi_t\psi_{t\rho}
        -2\rho\psi_\rho^3+2\psi_\rho^2
        -\rho\psi_\rho\psi_{\rho\rho}+\psi_{\rho\rho},
        \\[0.25em]
F_5={}&-2\psi_\rho\psi_t^2-\psi_t\psi_{t\rho}
        +2\psi_\rho^3+\psi_\rho\psi_{\rho\rho}.
\end{aligned}
\end{equation}

	  \bibliographystyle{hhieeetr}
	\bibliography{biblio}

\end{document}